\documentclass{article}

\usepackage{amssymb}
\usepackage{amsmath}
\usepackage[english]{babel}
\usepackage{graphicx}
\usepackage{textcomp}
\usepackage{cite}
\usepackage[linktocpage,unicode]{hyperref}

\def\s{{\,\rm s}}
\def\g{{\,\rm g}}
\def\eV{\,{\rm eV}}

\def\({\left(}
\def\){\right)}
\def\cm{{\,\rm cm}}

\def\beq{\begin{equation}}
\def\eeq{\end{equation}}
\def\bea{\begin{eqnarray}}
\def\eea{\end{eqnarray}}

\def\tz{\tilde{z}}

\begin{document}



\title{Signatures of primordial black hole dark matter}

\author{K.~M.~Belotsky$^{1}$\thanks{k-belotsky@yandex.ru}, A.~D.~Dmitriev$^{1}$, E.~A.~Esipova$^{1}$, V.~A.~Gani$^{1,2}$, A.~V.~Grobov$^{1}$,\\ M.~Yu.~Khlopov$^{1,3}$, A.~A.~Kirillov$^{1}$, S.~G.~Rubin$^{1}$, I.~V.~Svadkovsky$^{1}$\\[0.5cm]
	$^{1}$ National Research Nuclear University MEPhI\\
	(Moscow Engineering Physics Institute), Moscow, Russia\\
	$^{2}$ National Research Centre Kurchatov Institute,\\ Institute for Theoretical and Experimental Physics, Moscow, Russia\\
	$^{3}$ APC laboratory 10, rue Alice Domon et L\'eonie Duquet, 75205 Paris Cedex 13, France}















\date{}

\maketitle


\begin{abstract}
The nonbaryonic dark matter of the Universe is assumed to consist of new stable forms of matter. Their stability reflects symmetry of micro world and mechanisms of its symmetry breaking. In the early Universe heavy metastable particles can dominate, leaving primordial black holes (PBHs) after their decay, as well as the structure of particle symmetry breaking gives rise to cosmological phase transitions, from which massive black holes and/or their clusters can originate. PBHs can be formed in such transitions within a narrow interval of masses about $10^{17}$~g and, avoiding severe observational constraints on PBHs, can be a candidate for the dominant form of dark matter. PBHs in this range of mass can give solution of the problem of reionization in the Universe at the redshift $z\sim 5\ldots 10$. Clusters of massive PBHs can serve as a nonlinear seeds for galaxy formation, while PBHs evaporating in such clusters can provide an interesting interpretation for the observations of point-like gamma-ray sources. Analysis of possible PBH signatures represents a universal probe for super-high energy physics in the early Universe in studies of indirect effects of the dark matter.

\end{abstract}


\section{Introduction}

According to the modern cosmology, the dark matter, corresponding to $\sim 25\%$ of the total cosmological density, is nonbaryonic and consists of new stable forms of matter (see e.g.\cite{2013IJMPA..2830042K,2010RAA....10..495K} for review and references) that saturate the measured dark matter density and decouple from plasma and radiation at least before the beginning of the matter dominated (MD) stage. A huge landscape of possible candidates for the dark matter is proposed as well as their mixtures in multi-component dark matter scenarios are also possible \cite{2013IJMPA..2830042K,2010RAA....10..495K}.

Primordial black holes (PBHs) are a very sensitive cosmological probe for physical phenomena occurring in the early Universe. They could be formed by many different mechanisms (see e.g.~\cite{2013IJMPA..2830042K,2010RAA....10..495K} for review and references).

Being formed, PBHs should retain in the Universe and, if survive up to the present time, represent a specific form of the dark matter. Effect of PBH evaporation by S.W.Hawking \cite{1975CMaPh..43..199H} makes evaporating PBHs a source of fluxes of products of evaporation, particularly of $\gamma$ radiation \cite{1976ApJ...206....1P}. In a wide range of parameters the predicted effect of PBHs contradicts the data and it puts restrictions on mechanism of PBH formation and the underlying physics of very early Universe. On the other hand, at some fixed values of parameters, PBHs or effects of their evaporation can provide a nontrivial solution for astrophysical problems.

Here we outline some signatures of PBHs as a dominant or subdominant component of the dark matter. We briefly discuss the way for the PBH spectrum to reflect properties of superheavy metastable particles and parameters of phase transitions in the inflationary and post-inflationary stages (Section~\ref{hep}). These mechanisms of PBH formation have a specific feature of fixed range or even fixed values of PBH mass, what leads to effects of the PBH dark matter considered in Section~\ref{effects}.
The impact of constraints and cosmological scenarios, involving PBHs, on high energy physics is briefly discussed in Section~\ref{Discussion}.

\section{Mechanisms of PBH formation}
\label{hep}

\subsection{PBHs from superheavy metastable particles}

A black hole (BH) formation is highly improbable in homogeneous expanding Universe, since it implies metric fluctuations of order 1.
In the Universe with the equation of state $p=\gamma \epsilon,$ with numerical factor $\gamma$ being in the range $0 \le \gamma \le 1$, a probability to form BH from fluctuation within cosmological horizon is given by (see e.g. \cite{2010RAA....10..495K} for review and references)
\begin{equation}
	\label{ProbBH}
	W_\text{PBH}\propto \exp \left(-\frac{\gamma^2}{2
	\left\langle \delta^2 \right\rangle}\right).
\end{equation}
It provides exponential sensitivity of the PBH spectrum to softening of the equation of state in the early Universe ($\gamma \rightarrow 0$) or to increasing of the ultraviolet part of the density fluctuations spectrum ($\left\langle \delta^2 \right\rangle \rightarrow 1$). These phenomena can appear as cosmological consequence of the particle theory.

It was first noticed in \cite{1980PhLB...97..383K} that dominance of superheavy metastable particles with lifetime $\tau \ll 1\s$ in the Universe before their decay at $t \le \tau$ can result in formation of PBHs, retaining in the Universe after the particles decay and keeping some information on particle properties in their spectrum.

The mechanism \cite{2010RAA....10..495K, 1980PhLB...97..383K} is effective for formation of the PBHs with mass in the interval
\beq \label{Mint}
	M_0 \leq M \leq M_\text{BHmax}.
\eeq
The minimal mass corresponds to the mass within cosmological horizon in the period $t\sim t_0,$ when particles with mass $m$ and relative abundance $r=n/n_r$ (where $n$ is frozen out concentration of particles and $n_r$ is concentration of the relativistic species) start to dominate in the Universe. This mass is equal to \cite{2010RAA....10..495K, 1980PhLB...97..383K}
\begin{equation}
	\label{MBHmin}
	M_{0} = \frac{4 \pi}{3} \rho t^3_0 \approx m_\text{Pl}\left(\frac{m_\text{Pl}}{r m}\right)^2.
\end{equation}
The maximal mass is indirectly determined by the condition that fluctuation in the considered scale $M_\text{BHmax}$, entering the horizon at $t(M_\text{BHmax})$ with an amplitude $\delta(M_\text{BHmax})$ can manage to grow up to nonlinear stage, decouple and collapse before particles decay at $t=\tau$. For the scale invariant spectrum $\delta(M)=\delta_0$ the maximal mass is given by \cite{2010RAA....10..495K}
\begin{equation}
	\label{MBHmax}
	M_\text{BHmax} = m_\text{Pl} \frac{\tau}{t_\text{Pl}} \delta_0^{-3/2} =m_\text{Pl}^2 \tau \delta_0^{-3/2}.
\end{equation}

\subsection{Spikes from phase transitions in the inflationary stage}

Scale non-invariant spectrum of fluctuations, in which amplitude of small scale fluctuations is enhanced, can be another factor increasing the probability of PBH formation. The realistic treatment of the existence of other scalar fields along with inflaton in the period of inflation can give rise to the spectra with distinguished scales, determined by parameters of the considered fields and their interaction. In chaotic inflation scenario interaction of a Higgs field with inflaton can give rise to phase transitions in the inflationary stage. Such phase transitions in the inflationary stage lead to the appearance of specific spikes in the spectrum of the initial density perturbations. These spike-like perturbations, on scales that cross the horizon (60--1) $e$-- folds before the end of the inflation reenter the horizon during the radiation or the dust-like era and could in principle collapse to form PBHs. The possibility of such spikes in chaotic inflation scenario was first pointed out in \cite{1987NuPhB.282..555K} and implemented in \cite{1993PAN....56..412S} as a mechanism of PBH formation.

If phase transition takes place at $e$--folding $N$ before the end of inflation, the spike re-enters horizon in the radiation dominance (RD) stage and forms BH of mass
\beq
	\label{Mrd}
	M \approx \frac{m^2_\text{Pl}}{H_0} \exp(2 N),
\eeq
where $H_0$ is the Hubble constant in the period of inflation.

If the spike re-enters horizon in the MD stage it should form BHs of mass
\beq \label{Mmd}
M \approx
\frac{m^2_\text{Pl}}{H_0} \exp(3 N).
\eeq

\subsection{PBHs from the first order phase transitions in the early Universe}\label{phasetransitions}

First order phase transition goes through bubble nucleation, in which the false vacuum state decays leading to a nucleation of bubbles of the true vacuum and their subsequent expansion \cite{2010RAA....10..495K}. The potential energy of the false vacuum is converted into kinetic energy of the bubble walls thus making them highly relativistic in a short time. The bubble expands untill it collides with another one. As it was shown in \cite{1999PAN....62.1593K}, a BH may be created in a collision of two bubbles. The mass of this PBH is given by (see \cite{1999PAN....62.1593K} for details)
\begin{equation}
	\label{15}
	M_{\rm BH}=\gamma _1M_\text{bub},
\end{equation}
where $\gamma _1\simeq 10^{-2}$ and $M_{bub}$ is the mass that could be contained in the bubble volume at the epoch of collision in the condition of a full thermalization of bubbles.

If inflation ends by the first order phase transition, collision between bubbles of Hubble size in percolation regime leads to copious PBH formation with masses

\begin{equation}
	\label{16}M_0=\gamma _1M_\text{end}^\text{hor}= \frac{\gamma_1}2\frac{m_\text{Pl}^2}{H_\text{end}},
\end{equation}
where $M_\text{end}^\text{hor}$ is the mass of Hubble horizon at the end of inflation. According to \cite{1999PAN....62.1593K} the initial mass fraction of this PBHs is given by $\beta _0\approx\gamma _1/e\approx 6\cdot 10^{-3}$. For example, for typical value of $H_{\rm end}\approx 4\cdot 10^{-6}m_\text{Pl}$ the initial mass fraction $\beta $ is contained in PBHs with mass $M_0\approx 1$~g.

\subsection{Dark matter structures from extra dimensions}\label{extradimensions}

Extra space as a source of the dark matter attracts attention for a long time.  In this case WIMPs of various sorts are usually connected to the first KK excitation states, see e.g. \cite{bib:Cheng, bib:Belyaev}.

New idea concerning the essence of the dark matter and its origin was proposed in \cite{GaDmRu} in the framework of multidimensional gravity with higher derivatives. According to the authors' idea the geometry of compact extra dimensions depends on the initial conditions. It was shown there that space domains of the dark matter could appear if geometry of a 2-dim extra space inside the domains and outside them is different. Such regions contain some excess of the energy density, interact only gravitationally and hence may be considered as a dark matter candidate.

These domains move in the Galaxy with the virial velocity about $300$ km/s and represent the cold dark matter. It is interesting problem to find a method of their detection because ordinary particle detectors are not aimed for such events. Indeed, as was shown in \cite{GaDmRu}  one can neglect the excitations of the domain during its non-relativistic interaction with nucleus of a detector provided the multidimensional Planck mass is about $m_D \sim 10^6$ GeV. In this case pure elastic scattering of nuclei on the domain as a whole takes place. The question is whether one can distinguish such a ``Moessbauer'' scattering and the ordinary one.

Minimal size of such objects is comparable to a size of the extra space with the mass varying in wide range  depending on the initial conditions and numerical values of the model parameters. More or less natural mass of WIMPs in this model is about 10 TeV.  In this case new methods of the dark matter detection has to be developed. The primary estimation of such objects interaction with nucleons leads to the cross section about $10^{-43}$~cm$^2$ what does not contradict constraints on very massive dark matter particles.


\subsection{PBHs from closed walls in the inflationary Universe}\label{walls}

A wide class of particle models possesses a symmetry breaking pattern, which can be effectively described by pseudo-Nambu-Goldstone field and which corresponds to formation of unstable topological defect structure in the early Universe (see \cite{2010RAA....10..495K} for review and references). The Nambu-Goldstone nature in such an effective description reflects the spontaneous breaking of global U(1) symmetry, resulting in continuous degeneracy of vacua. The explicit symmetry breaking at smaller energy scale changes this continuous degeneracy by discrete vacuum degeneracy. It gives rise to specific mechanisms of PBH formation.

If phase transitions take place in the inflational stage, new forms of primordial large scale correlations appear. The value of phase after the first phase transition at the energy scale $f$ is inflated over the region corresponding to the period of inflation, while fluctuations of this phase change in the course of inflation its initial value within the regions of smaller size. Owing to such fluctuations, for the fixed value of $\theta_{60}$ in the period of inflation with {\it e-folding} $N=60$ corresponding to the part of the Universe within the modern cosmological horizon, strong deviations from this value appear at smaller scales, corresponding to later periods of inflation with $N < 60$. If $\theta_{60} < \pi$, the fluctuations can move the value of $\theta_{N}$ to $\theta_{N} > \pi$ in some regions of the Universe. After reheating in the result of the second phase transition at the scale $\Lambda \ll f$ these regions correspond to vacuum with $\theta_{vac} = 2\pi$, being surrounded by the bulk of the volume with vacuum $\theta_{vac} = 0$. As a result massive walls are formed at the border between the two vacua. Since regions with $\theta_{vac} = 2\pi$ are confined, the domain walls are closed. After their size equals the horizon, closed walls can collapse into BHs.

This mechanism can lead to formation of PBHs of a whatever large mass (up to the mass of active galactic nuclei (AGNs) \cite{2001JETP...92..921R}, see for latest review \cite{2010RAA....10..495K}). Such BHs appear in the form of the PBH clusters, exhibiting fractal distribution in space. It can shed new light on the problem of galaxy formation \cite{2010RAA....10..495K, 2001JETP...92..921R}.

The mass range of formed BHs is constrained by fundamental parameters $f$ and $\Lambda$ of the model. The maximal BH mass is determined by the condition that the wall does not dominate locally before it enters the cosmological horizon. Otherwise, local wall dominance leads to a superluminal $a \propto t^2$ expansion for the corresponding region, separating it from the other part of the Universe. This condition corresponds to the mass \cite{2010RAA....10..495K}
\beq
\label{Mmax} M_{max} =
\frac{m_{\rm Pl}}{f}m_{\rm Pl}\left(\frac{m_{\rm Pl}}{\Lambda}\right)^2.
\eeq
The minimal mass follows from the condition that the gravitational radius of BH exceeds the width of the wall and it is equal to \cite{2010RAA....10..495K,2001JETP...92..921R}
\beq
\label{Mmin} M_{min} = f\left(\frac{m_{\rm Pl}}{\Lambda}\right)^2.
\eeq

\subsection{$10^{17}\g$ PBHs from the closed walls}

The mechanism of PBH formation from the closed walls can be implemented for creation of PBH with mass $\sim 10^{17}$ g. PBHs with similar mass avoid constraints from observations by gamma-radiation data and femto/micro lensing effects \cite{2010PhRvD..81j4019C}, so they can provide the dominant part of the dark matter. Attempt to constrain PBH with mass $>10^{17}$ g, based on a tidal capture of PBH by neutron star \cite{Loeb}, met some counter-evidence \cite{Tinyakov, Tinyakov2}.

\begin{figure}[t]
\centering
\includegraphics[width=0.4\textwidth]{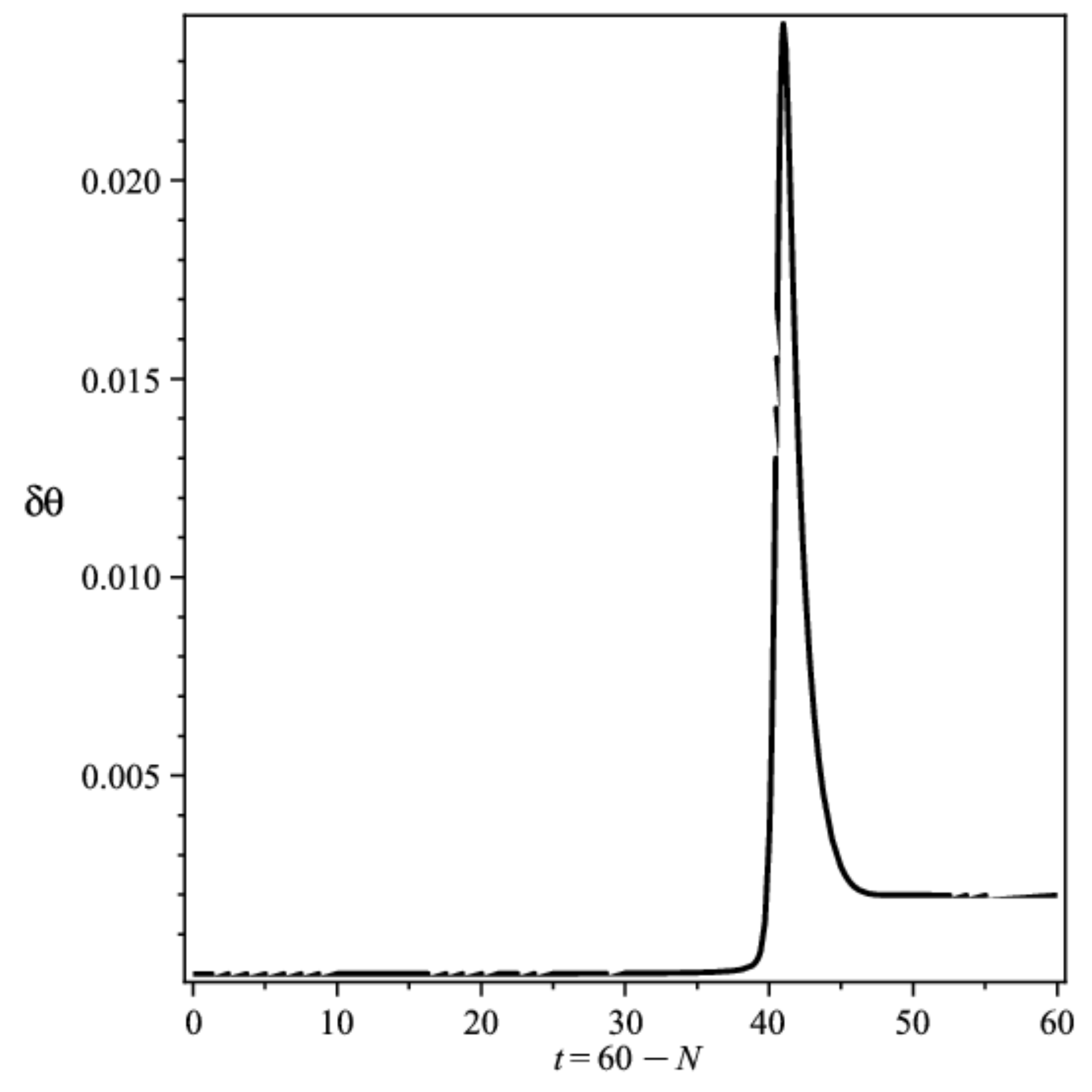}
\caption{Computed value of $\delta \theta$ as a function of (60-N), where $N$ is the number of the e-fold, $N=H\cdot t$;}
\label{dTheta}
\end{figure}

Here we follow to the specific mechanism developed in \cite{2001JETP...92..921R, 2000hep.ph....5271R, 2011GrCo...17..181G}. The basis of this mechanism is quantum fluctuations of scalar field with a potential possessing at least two minima.

\begin{figure}[t]
\centering
\includegraphics[width=0.4\textwidth]{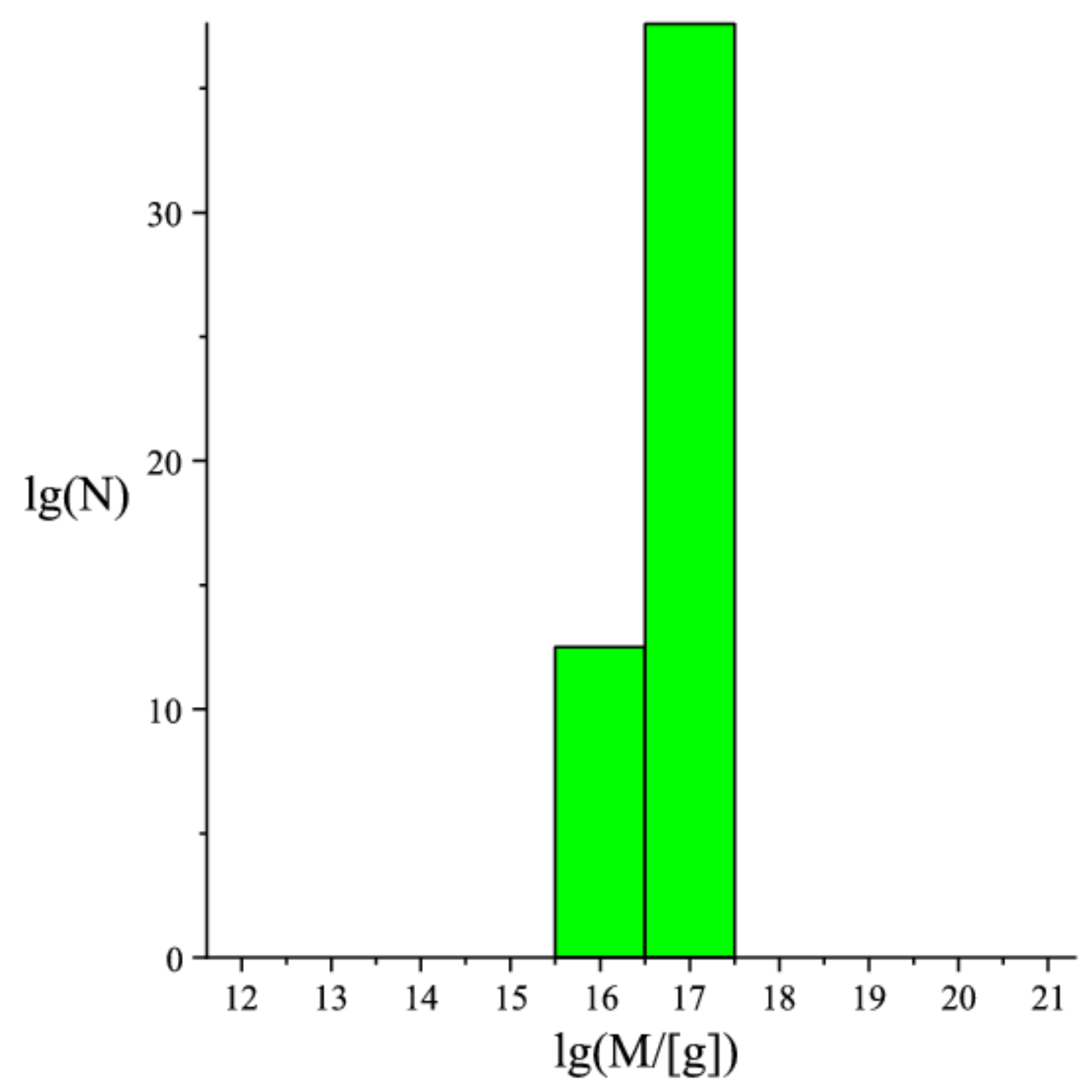}
\caption{Histogram of a PBH mass spectrum, under the parameters we use: $f=80H$, $\lambda=8 \cdot 10^{-5}$, $C=4 \cdot 10^3 H^{a}$, $a=1.455$, $\chi_0 = 4.4 \cdot 10^4 H$, $\theta_0 = 1.94$, $\Lambda=0.6 H$}
\label{spectrum PBH}
\end{figure}

Let there be a scalar field with the potential
\begin{align} \label{potential}
V_0(\Psi)=\lambda\left(|\Psi|^2-\frac{f^2}{2}\right)^2,
\end{align}
where $\Psi = \chi(t)\exp(i\varphi/f)$, $\chi(t)$ is a radial component of $\Psi$ field with minimum at $f/\sqrt{2}$, $\varphi$ is a massless Nambu-Goldstone field.  Landscape ideas \cite{Landscape, random, random1, random2} deal with potentials of arbitrary form. Here we propose specific form of the potential
\begin{align} \label{potential2}
&V(\chi)=V_0(\chi) \cdot F(\chi), \\
F(\chi) = &\frac{C^2}{(|\chi|^a+C)^2}, \qquad \quad C, a = \text{const}>0  \nonumber
\end{align}
suitable for our purposes. The function $F(\chi)\xrightarrow[\chi \rightarrow 0]{} 1$, so that $V(\chi)\xrightarrow[\chi \rightarrow 0]{} V_0(\chi)$ and the modified potential is close to the original one while $\chi$ is small.
The presence of quantum corrections is responsible for the next term appearing in the potential:
\begin{align} \label{potential2}
\Lambda^4(1-\cos\theta),
\end{align}
so that $\Lambda$ is supposed to be small.
During inflation, when the friction term is large the classical motion of the angular field $\theta$ is frozen due to the smallness of the potential tilt. The dynamics of the radial field $\chi$ is governed by the equation of motion:
\begin{equation}\label{classeq}
\frac{\mathrm{d}^2 \chi(t)}{\mathrm{d} t^2}+3H\frac{\mathrm{d} \chi(t)}{\mathrm{d} t}+\frac{\mathrm{d} V\left(\chi\right)}{\mathrm{d} \chi} = 0,
\end{equation}
where $H$ is Hubble's parameter. This differential equation can be solved numerically.

While field $\chi$ is moving classically to the potential minimum $\chi\simeq f/\sqrt{2}$, the phase $\theta=\varphi/f$ changes due to the quantum fluctuations at the de Sitter background.
During inflation the amplitude of the phase $\theta$ changes as
\begin{equation}
\delta \theta \simeq \frac{H}{2\pi \cdot \overline{|\chi(t)|}}
\end{equation}
each e-fold, where  $\overline{|\chi(t)|}$ is the average absolute value of the field $\chi$ during the e-fold.

Discrete spectrum $N_{\rm PBH}(M)$ of specific masses $M$ of PBH can be easily obtained by iterative procedure \cite{2000hep.ph....5271R}.

Contribution of PBH to the average energy density  $\Omega_{\rm PBH}$ may be estimated as follows:
\begin{equation}\label{concentration}
   \Omega_{\rm PBH} = \sum_i \frac{M_{i}\cdot N_i}{\rho_c \upsilon_U} \approx 0.22,
\end{equation}
where the uniform space distribution of PBH is supposed.
Here $N_{i}$ is the number of PBHs of mass $M_i$, $\rho_c = 1.29 \cdot 10^{-29} h^2$ $\frac{\g}{\cm^3}$ is the critical density and $\upsilon_U\approx l^{3}_U$ is the volume of the Universe with $l_U \approx 6000$ Mpc.

\section{Effects of the PBH dark matter}\label{effects}

\subsection{PBH dark matter and reionization of the Universe}

As was noted, PBH with mass around $M_{17}=10^{17}$ g 
can provide the dominant form of the dark matter. 
The PBHs of these range of mass have even more attractive features: with the help of them one could explain positron line from the Galactic center \cite{INTEGRAL} due to effects of accretion \cite{Titarchuk} or the Hawking evaporation \cite{Dolgov}. Here we discuss possibility to explain with them a reionization of the Universe on the base of the Hawking evaporation effect.

In our estimations we take the mass range $10^{16}\g<M<10^{17}\g$ and abundance according to the upper limit \cite{2010PhRvD..81j4019C}, which can be represented in the form
\begin{equation}
\Omega_{\rm PBH}=\left\{
\begin{array}{l}
0.25,\text{ if $M>M_{\rm peak}$} \\
0.25\left(\frac{M}{M_{\rm peak}}\right)^{3.36},\text{ if $M<M_{\rm peak}$,}
\end{array}
\right.
\label{OmegaPBH}
\end{equation}
where $M_{\rm peak}=0.78\cdot 10^{17}$ g.
The evaporation temperature for such PBH is $T_{\rm ev}\approx 0.1\frac{M_{17}}{M}$ MeV, the mean energy of evaporating photons is $\approx 6T_{\rm ev}$ and electrons and neutrinos is $\approx 4T_{\rm ev}$ \cite{2010PhRvD..81j4019C}. One can see how the temperature of baryonic matter and its ionization degree change being exposed to the Hawking radiation from PBHs.

\begin{figure}
	\centering
	\includegraphics[width=0.6\textwidth]{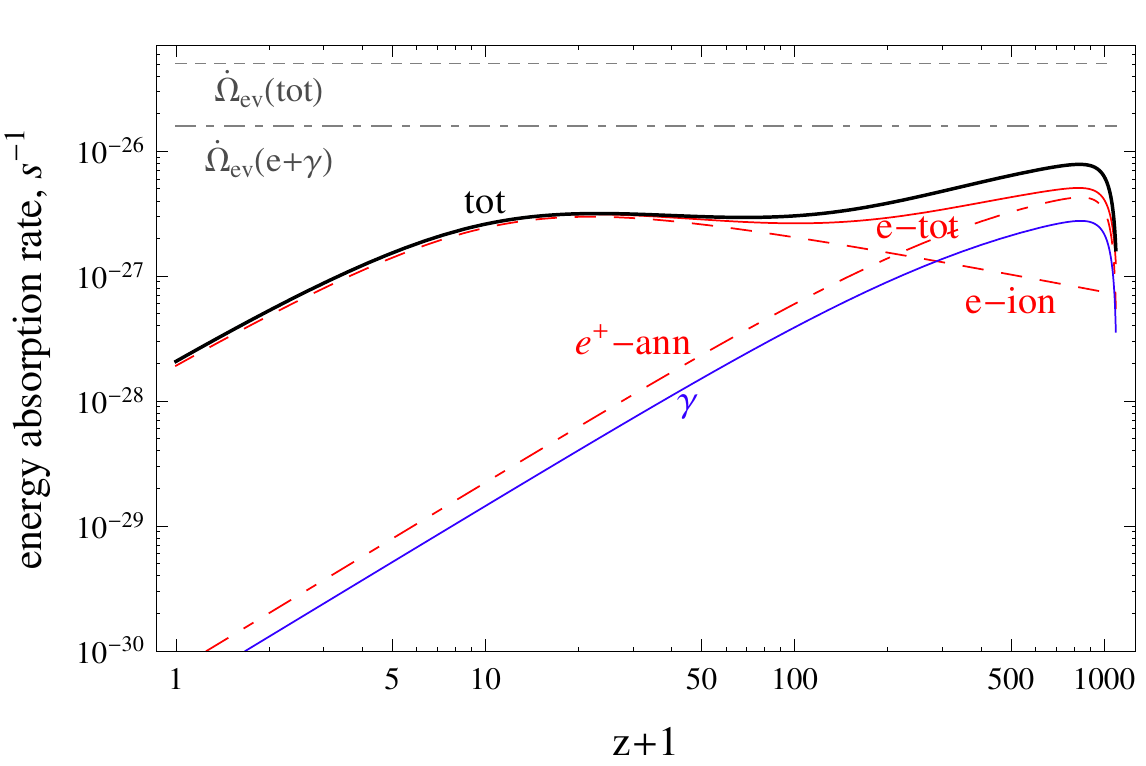}
	\caption{Energy absorption rates for all processes considered: $\dot{\Omega}_{\rm abs}^{(e{\rm-ion})}$, $\dot{\Omega}_{\rm abs}^{(e{\rm-ann})}$ and their sum, $\dot{\Omega}_{\rm abs}^{(\gamma)}$ and the total rate, for $M=M_{\rm peak}$. $\dot{\Omega}_{\rm ev}$ is also shown which illustrates total evaporation rate.}
	\label{rates}
\end{figure}

Let $\dot{\Omega}_{\rm ev}$ and $\dot{\Omega}_{\rm abs}$ be the rates of energy evaporation and absorption by the matter in units of the critical density. The first one can be approximated by
\begin{equation}
\dot{\Omega}_{\rm ev}= 
\frac{1}{3}\(\frac{M_U}{M}\)^3\frac{\Omega_{\rm PBH}(M)}{t_U}
\end{equation}
for $M\gg M_U$, where $M_U\approx 0.5\cdot 10^{15}$ g, $t_U\approx 14$ Gyr.

In the considered interval of evaporation temperature, PBH emits gravitons, photons, three sorts of neutrinos and electrons and positrons. The fractions of $e^{\pm}$ and photons in the evaporation flux can be estimated as $\kappa_{e^{\pm}}=\frac{0.57\hat{g}_e(M)}{1.02+0.57\hat{g}_e(M)}$ and $\kappa_{\gamma}=\frac{0.12}{1.02+0.57\hat{g}_e(M)}$ respectively, where $\hat{g}_e(M)$ is the suppression factor when $T\lesssim m_e$.

A photon from evaporation in the energy range of question ($\omega\sim 0.5\dots 5$ MeV) loses its energy due to Compton scattering (CS) and red shift. CS energy loss rate can be characterized by the respective inverse time $\tau_C^{-1}=n_H\sigma \frac{\Delta \omega}{\omega} c = t_U\tz_C^{3/2}\tz^{-3}$, where $\sigma$ is the Klein-Nishina cross section, $\Delta\omega$ is the energy transfer in one scattering,  $n_H=1.9\cdot 10^{-7}\cm^{-3}\tz^3$ is the total number density of hydrogen. Denotation $\tz_{(i)}\equiv z_{(i)}+1$ is introduced, MD stage is taken into account only. In our calculation, numeric factor $\tz_C$ is assumed to be independent on $\omega$ and equal to $\tz_C\approx 340$ (see details in \cite{reion}). Redshift rate is given by Hubble parameter $H=2/3\,t_U^{-1}\tz^{3/2}$.

Since Compton scattering only provides energy transfer to baryonic matter, respective rate can be roughly estimated through the ratio of the rates of energy losses
\begin{equation}
\dot{\Omega}_{\rm abs}^{(\gamma)}(z)=\kappa_{\gamma}\dot{\Omega}_{\rm ev}\frac{\tau_C^{-1}}{\tau_C^{-1}+c_HH}=
\kappa_{\gamma}\dot{\Omega}_{\rm ev}\frac{\tz^{3/2}}{\tz^{3/2}+\frac{2c_H}{3}\tz_C^{3/2}}.
\label{RGamma}
\end{equation}
Here we include factor $c_H\approx 3$ to adjust this approximation to more accurate one \cite{reion}.

Electrons and positrons from evaporation of PBH should experience energy losses due to scattering off the cosmic microwave background (CMB) photons, ionization and red shift. Effects of interaction with plasma which has a very low density at the considered redshift are not discussed here.


Energy losses on the CMB in the ultra-relativistic limit are given by \cite{Ginzburg} $\left.\frac{dE}{dt}\right|_{\rm CMB}=-\beta E^2$, where $\beta =\omega_2^{-1} t_U^{-1}\tz^4$ is defined by the CMB energy density, $\omega_2\approx 90$ MeV. Note that each scattering leads to energy transfer as small as $\sim (E/m)^2$ of the primary CMB photon energy. The energy loss time scale is given by $\tau_{\rm CMB}=(\beta E)^{-1}$. The ionization losses rate will be approximated by $dE/dx=\text{const} \approx 4$ MeV g$^{-1}$cm$^2$. Characteristic time of the ionization losses is defined as $\tau_{\rm ion}=E/(dE/dt)_{\rm ion}\approx \frac{E}{\omega_1}t_U\tz^{-3}$, where $\omega_1=0.016$ MeV. Losses on CMB is much larger than ionization losses over the most of period of interest. In the late period ($z\lesssim 10$) the ionization rate approaches the CMB one, but both become comparable with the expansion rate.

\begin{figure}[t]
	\centering
	\includegraphics[width=0.6\textwidth]{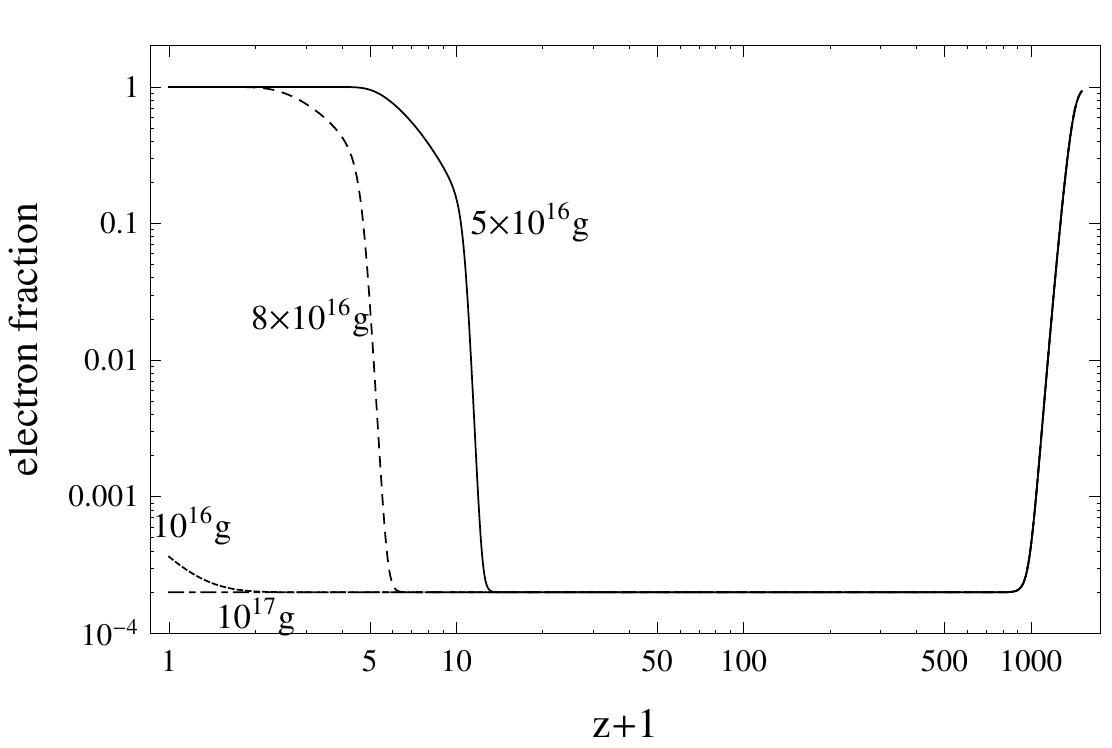}
	\caption{Electron fraction $x_e$ dependence on the redshift.}
	\label{xe}
\end{figure}

Energy absorption rate is defined here by the ionization process, so we can estimate it roughly as
\begin{equation}
\dot{\Omega}_{\rm abs}^{(e\text{-ion})}=\kappa_e\dot{\Omega}_{\rm ev}\frac{\tau_{\rm ion}^{-1}}{\tau_{\rm ion}^{-1}+\tau_{\rm CMB}^{-1}+c_HH}=\kappa_e\dot{\Omega}_{\rm ev}\frac{\frac{\omega_1}{E}\tz^{3/2}}{\frac{\omega_1}{E}\tz^{3/2}+\frac{E}{\omega_2}\tz^{5/2}+\frac{2c_H}{3}},
\label{Rion}
\end{equation}
where $E=4T_{\rm ev}$.

One more contribution into heat of matter should be given by the annihilation of the stopped positrons. Part of the energy release rate, $\frac{m_e}{E}\kappa_e\dot{\Omega}_{\rm ev}$, comes into annihilation photons, which absorption is described by Eq.\eqref{RGamma}. So,
\begin{equation}
\dot{\Omega}_{\rm abs}^{(e\text{-ann})}=\frac{m_e}{E}\kappa_e\dot{\Omega}_{\rm ev}\frac{\tz^{3/2}}{\tz^{3/2}+\frac{2c_H}{3}\tz_C^{3/2}}.
\label{Rann}
\end{equation}
Numerical results given by Eqs.\eqref{RGamma}, \eqref{Rion}, \eqref{Rann} are within factor two of more accurate approximation \cite{reion}. Below we shall present the results, obtained with more accurate approximation, which are qualitatively described by Eqs.\eqref{RGamma}--\eqref{Rann}.

All absorption rates for $M=M_{\rm peak}$ are shown in Fig.~\ref{rates}. It is seen that the electron-positron ionization losses, which are suppressed by losses in the CMB scattering and at $z\lesssim 10$ by red shift also, are found nonetheless to be the most effective mechanism of the energy transfer from the PBH radiation to matter. Total energy absorbed by baryonic matter by the moment $z\sim 5\ldots 10$, makes up $1\ldots 2$ eV per each atom. Let us now consider temperature arguments.

Evolution of the temperature of hydrogen can be described by the equation \cite{reion}
\begin{equation}
\frac{dT}{dt}=\frac{2\dot{\Omega}_{\rm abs}m_p}{3\cdot0.76\Omega_B(1+x_e)}-\frac{8\pi^2}{45}T_{\gamma}^4\sigma_T\frac{x_e}{1+x_e}\frac{T-T_{\gamma}}{m_e}-2HT.
\label{dT}
\end{equation}
Here $\sigma_T$ is the Thomson cross section, 
$x_e$ is the ratio of free electron number density and $n_H$. The latter is defined from the Saha formula
\begin{equation}
\frac{x_e^2}{1+x_e}=\frac{1}{n_H}\(\frac{m_eT}{2\pi}\)^{3/2}\exp\(-\frac{13.6 \eV}{T}\).
\label{Saha}
\end{equation}
Note that the second term in the right side of Eq.~\eqref{dT} takes into account the CMB-matter(electrons) energy exchange. 
Solving Eq.~\eqref{dT} gives $z$-dependence of the electron fraction $x_e$ for different PBH masses as shown in Fig.~\ref{xe}. Minimal $x_e$ was formally taken to be $2\cdot 10^{-4}$, corresponding to its frozen magnitude.
%

\begin{figure}[t]
	\centering
	\includegraphics[width=0.6\textwidth]{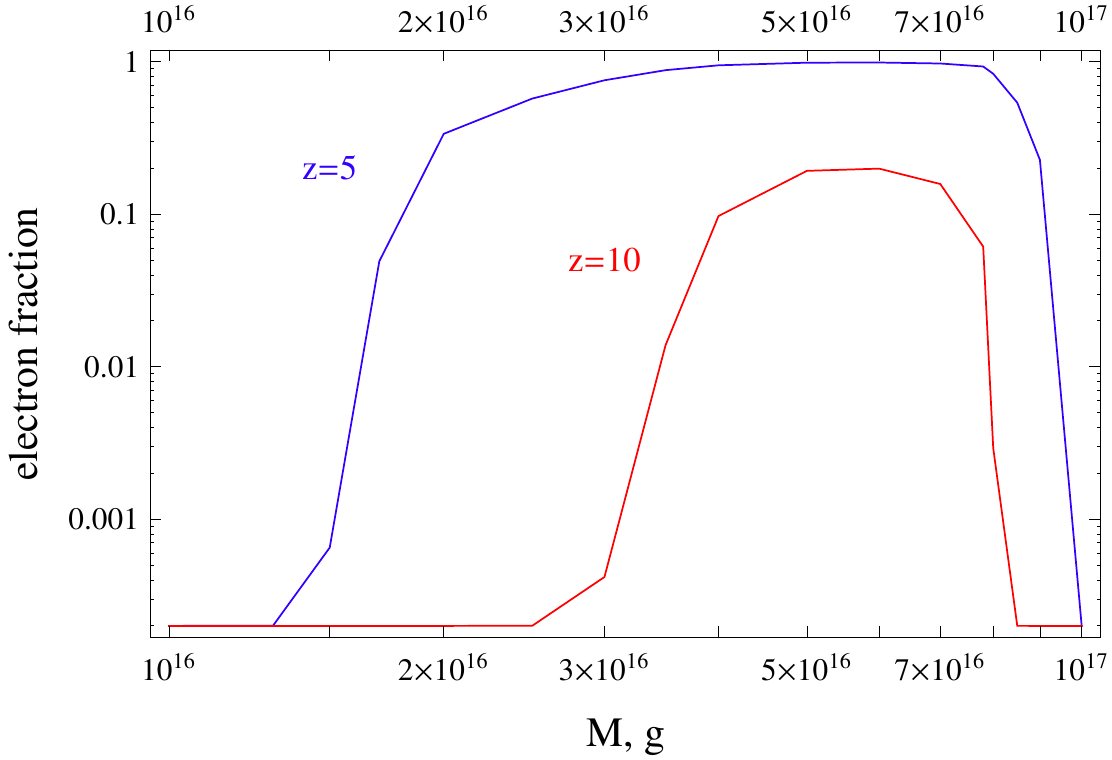}
	\caption{Electron fraction $x_e$ dependence on the PBH mass.}
	\label{xeM}
\end{figure}

%
Figure \ref{xeM} shows $x_e$ as a function of the PBH mass $M$ for $\tz=5$ and 10. 
As it is seen, PBH with $M=3\cdot 10^{16}\dots 8\cdot 10^{16}$ g could provide reionization of the Universe.

\subsection{Primordial seeds for AGNs}

In \cite{2003A&AT...22..727D} possible consequences of the existence of pregalactic population of BHs with masses $M \sim 10^{5} M_{\odot}$ before the recombination time were discussed. These hypothetical PBHs are supposed to be mixed with dark matter because of their cosmological origin. As a result the total mass of these PBHs in any galaxy would be proportional to the galactic dark matter halo mass. And according to the model \cite{2003A&AT...22..727D} it leads to the correlations between central BHs masses $M_{BH}$ and galactic bulge velocity dispersions $\sigma_e$ in the form $M_{BH} \propto \sigma_{e}^{4}$, as observed. Thus, such form of correlation supports the idea of pregalactic origin of the massive BHs as the AGN seeds.

In \cite{2008ARep...52..779D, 2005GrCo...11...99D} the very early galaxy and quasar formation is discussed in the framework of the model based on PBH cluster as a source of the initial density perturbation.

The paper \cite{2001JETP...92..921R} proposes another mechanism of protogalaxies formation. It is based on the second-order phase transition at the inflationary stage. It is shown that collapsing closed domain walls could lead to the massive BH clusters formation, which in turn can serve as nuclei for the future galaxies.

\subsection{Clusters of BHs as point-like $\gamma$-ray sources}

\begin{figure}[t]
  \centering
  \includegraphics[width=0.6\textwidth]{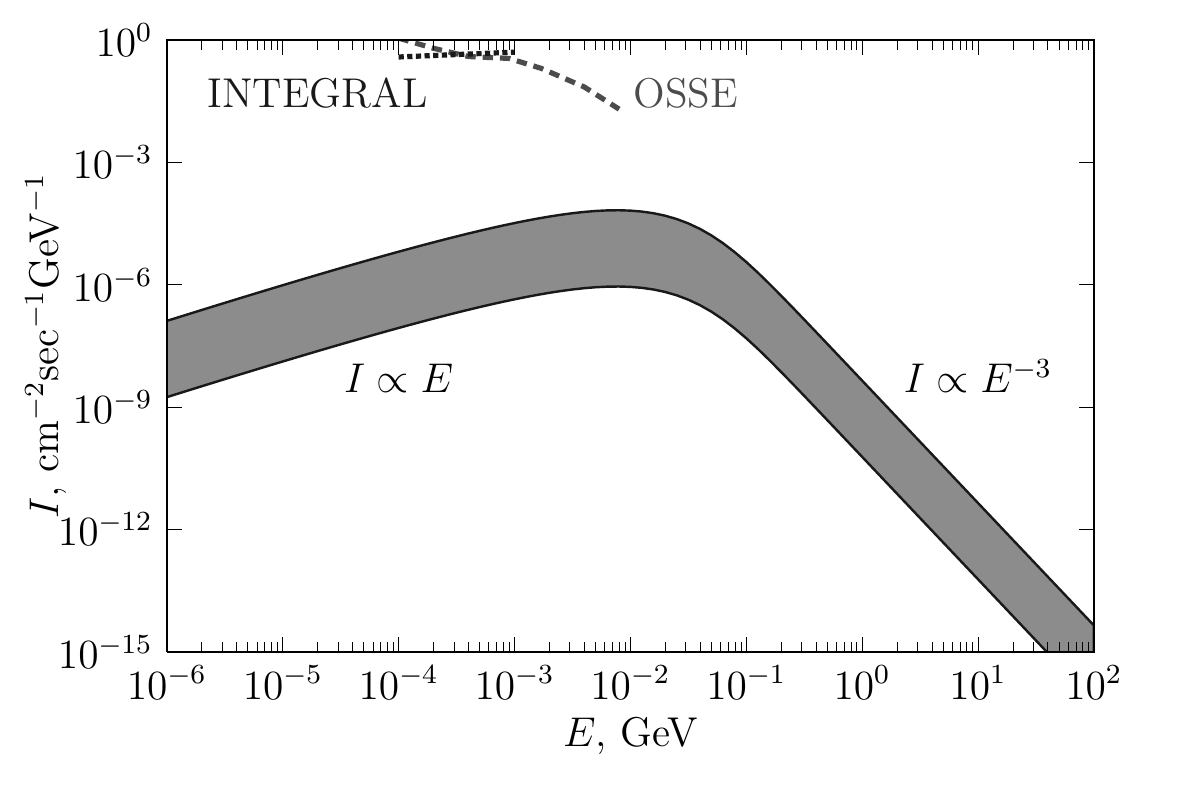}
  \caption{Expected spectrum of photons from the PBH cluster (filled area). Below this area, the flux becomes lower than the LAT 
  sensitivity ($r>R_\text{max})$. Differential sensitivities of X-ray detectors INTEGRAL and OSSE are presented as well.}
  \label{fig:BHMassSpectrum}
\end{figure}

Due to weak accretion of surrounding matter, observation of BHs with masses less than $10^{4}M_{\odot}$ can be quite difficult. An alternative method for the BHs detection by means of the Hawking's radiation \cite{1975CMaPh..43..199H} is effective only for BHs with low masses that are situated near the Earth. From \cite{2011GrCo...17...27B, 2011APh....35...28B} one can conclude that situation can be different for PBH clusters with a large number of small BHs (with masses $\lesssim 10^{15}$~g). The integral Hawking's radiation of such a cluster could then be sufficient for detection by modern gamma-ray telescopes on the Earth.

The number of clusters and their properties depend on the choice of the scalar field potential parameters
\cite{2008ARep...52..779D}
\begin{equation}
	V= \lambda \left(\phi^{*}\phi-\frac{1}{2}f^{2}\right)^{2} +\Lambda^{4}\left(1-\cos\theta\right).
\end{equation}
With the choice of parameters $f = 10^{14}$ GeV, $\Lambda = 1.66\cdot10^{13}$ GeV and almost arbitrary value of $\lambda$, one can obtain cluster structures with the mass distribution, approximately described by the power law dependence:
\begin{equation}
	\label{eq:BHMassSpect}
	f(M) = \cfrac{4.4\cdot 10^{17}}{M_U}\,
	 \left(\cfrac{M}{M_U}\right)^{2} \left(1+\left(\cfrac{M}{M_U}\right)^{3}\right)^{-4/3}.
\end{equation}
Here we take into account change in the spectrum caused by the evaporation of light BHs during the lifetime of a cluster. With this choice of parameters there are $N_\text{cl}\sim$1400 clusters with the mass $M\approx 9.5 M_\odot$ and of characteristic sizes $R\sim$1~pc in a galactic halo. Note that the abundance of such clusters ($\Omega_\text{PBH}\sim 3\cdot 10^{-10}$) doesn't contradict to the modern constraints on the PBH density \cite{2010PhRvD..81j4019C}.

Due to the Hawking's mechanism clusters act as permanent sources of cosmic rays, neutrino and gamma radiation. Let us estimate a typical brightness of the PBH cluster in the gamma range.
The total flux 
of photons from the cluster of PBHs, radiating with intensity $\frac{dN_\gamma}{dE\,dt}$, is given by the integral
(we take into account photon energies   $E_{\gamma}>E_1=100$~MeV, corresponding to the threshold energy of the Fermi LAT gamma ray telescope)
\begin{equation}
	\dot{N}=\int\limits_{m_\text{Pl}}^{M_\text{max}}f(M)\,dM
	\int\limits_{E_1}^{\infty}\frac{dN_\gamma}{dE\,dt}
	\left(E,M\right)\,dE
	\approx 6.6\cdot 10^{36}\;\text{sec}^{-1}.
\end{equation}

For the sensitivity of the Fermi LAT telescope to a point $\gamma$-source $F_\text{min}=3\cdot 10^{-9}$~cm$^{-2}$sec$^{-1}$ \cite{2009ApJ...697.1071A}, the maximal distance $R_\text{max}$ at which the clusters could be registered by a detector as $\gamma$-source is about $R_\text{max}\sim \sqrt{\dot{N}/F_\text{min}}\sim 4$~kpc. Then the number of such sources can be $ N\sim n_\text{cl} R_\text{max}^{3}\sim 30$, where $n_\text{cl}$ is the number density of clusters. 
%
Fig.~\ref{fig:BHMassSpectrum} represents the spectrum from the cluster at the distances $n_{\rm cl}^{-3}\ldots R_{\rm max}$. Note that due to low sensitivity of x-ray telescopes, detecting PBH clusters is beyond their abilities \cite{2003A&A...411L...1W,1992NASACP3137....3C}.

\begin{figure}[t]
  \centering
  \includegraphics[width=0.6\textwidth]{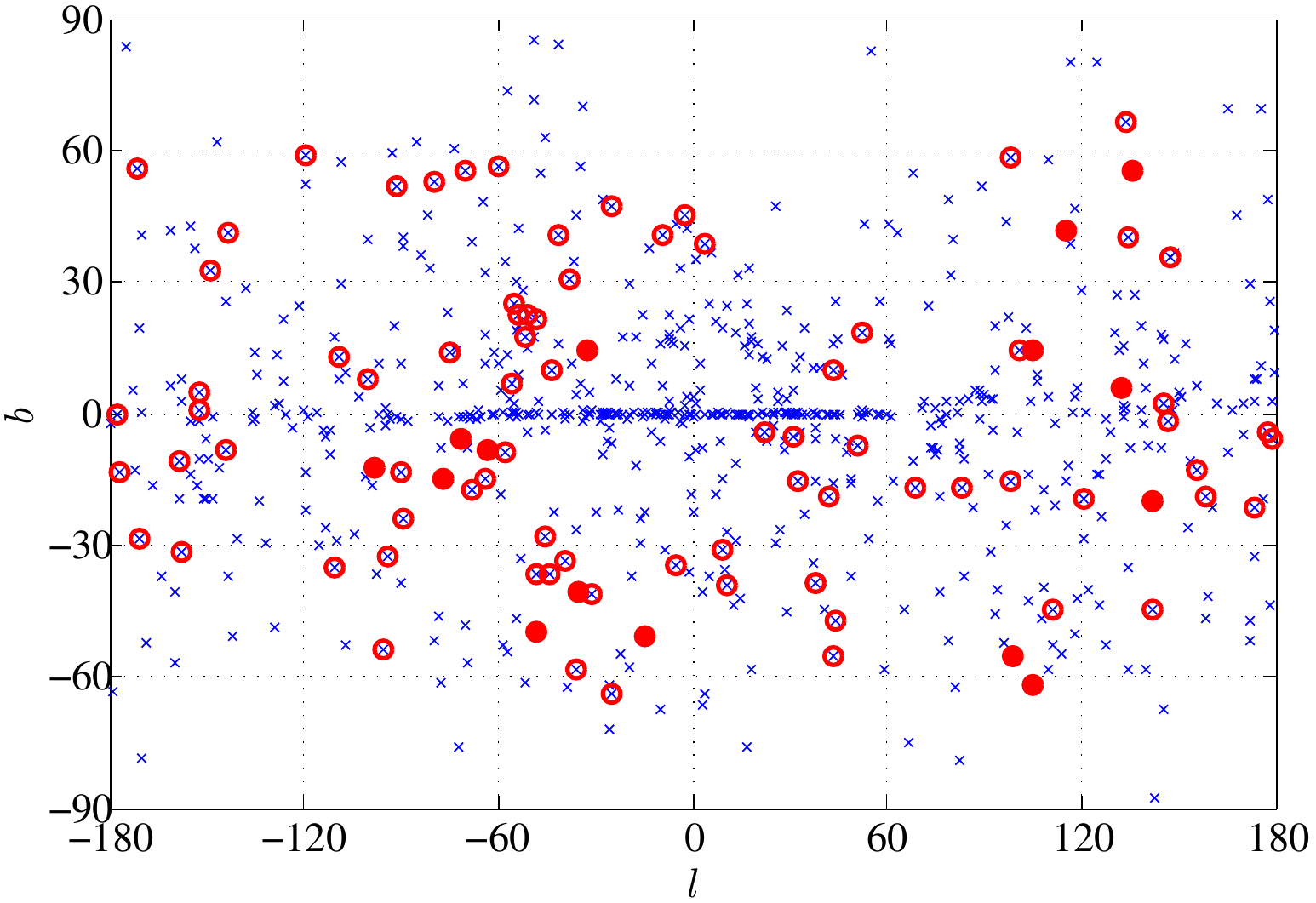}
  \caption{The distribution of unidentified gamma ray sources seen by Fermi LAT (crosses) is presented. Filled circles represent sources with the spectral index 3 (candidates to PBH clusters) within the error of 1$\sigma$, empty circles~--- sources with the spectral index 3 within the error of 3$\sigma$}
  \label{fig:BHCSph}
\end{figure}

Fermi LAT has discovered 15 sources with the spectral index 3, as PBH cluster featured, within the error of $1\sigma$ \cite{2010ApJS..188..405A}. These sources are uniformly distributed throughout the celestial sphere (see Fig.~\ref{fig:BHCSph}), and their number agrees well with the prediction obtained.

\section{Conclusions}\label{Discussion}

Being formed in the very early Universe as initially nonrelativistic form of matter, PBHs should have increased their contribution to the total density during the RD stage of expansion, while effect of the PBH evaporation should have strongly increased the sensitivity of astrophysical data to their presence. It links the hypothetical sources of cosmic rays or gamma background to parameters of superheavy particles in the early Universe and of the first and second order phase transitions, thus making a sensitive astrophysical probe to particle symmetry structure and pattern of its breaking at superhigh energy scales as well as to effects of extra dimensions. The possibility of PBH formation in definite ranges or even fixed values of the PBH masses add interesting aspects for such probes.

The presented approach sheds new light on the indirect effects of the dark matter. Being elusive for direct experimental searches of the dark matter particles, the indirect probe for the PBH dark matter hypothesis is of special interest.

\section*{Acknowledgments}

The work was supported by the Ministry of Education and Science of the Russian Federation, Project 3.472.2014/K, the work of K.M.B., A.V.G., M.Yu.Kh. and A.A.K. on PBH form of dark matter was supported by Grant RFBR 14-22-03048, the work of A.D.D. and I.V.S. was supported by Grant RFBR 14-02-31417.  The work of V.~A.~Gani was also supported by the Russian Federation Government under Grant No.~NSh-3830.2014.2.



\begin{thebibliography}{99}
\bibitem{2013IJMPA..2830042K}
M.~Y. Khlopov, {\em Int. J. Mod. Phys.~A} {\bfseries 28} 30042 (2013),  {\ttfamily arXiv: 1311.2468 [astro-ph.CO]}.

\bibitem{2010RAA....10..495K}
M.~Y. Khlopov, {\em RAA} {\bfseries 10} 495--528 (2010), {\ttfamily arXiv: astro-ph/0801.0116}.

\bibitem{1975CMaPh..43..199H}
S.~W. Hawking, {\em Comm. Math. Phys.} {\bfseries 43} (1975) 199--220.

\bibitem{1976ApJ...206....1P}
D.~N. Page and S.~W. Hawking, {\em Astrophys. J.} {\bfseries 206} (1976) 1--7.

\bibitem{1980PhLB...97..383K}
M.~Y. Khlopov and A.~G. Polnarev, {\em Phys. Lett.~B}  {\bfseries 97} (1980) 383--387.

\bibitem{1987NuPhB.282..555K}
L.~A. Kofman and A.~D. Linde, {\em Nucl. Phys.} {\bfseries B282} (1987) 555.

\bibitem{1993PAN....56..412S}
A.~S. Sakharov and M.~Y. Khlopov,  {\em Phys. of Atom.
  Nucl.} {\bfseries 56} (1993) 412--417.

\bibitem{1999PAN....62.1593K}
R.~V. Konoplich, {\em et~al.}, {\em Phys. of Atom. Nucl.} {\bfseries 62} (1999) 1593--1600.

\bibitem{bib:Cheng}
H.-C. Cheng, {\em et~al.}, {\em Phys. Rev.} {\bfseries D66} (2002) 036005, {\ttfamily arXiv: hep-ph/0204342}.

\bibitem{bib:Belyaev}
A. Belyaev, {\em et~al.}, {\em JHEP} {\bf 1306} (2013) 080, {\ttfamily arXiv: 1212.4858}.

\bibitem{GaDmRu}
V.~A.~Gani, A.~E.~Dmitriev, and S.~G.~Rubin, in preparation.

\bibitem{2001JETP...92..921R}
S.~G. Rubin, A.~S. Sakharov, and M.~Y. Khlopov, {\em Sov. Phys. JETP} {\bfseries 92} (2001) 921--929, {\ttfamily
  arXiv: hep-ph/0106187}.

\bibitem{2010PhRvD..81j4019C}
B.~J. Carr, {\em et~al.}, {\em Phys. Rev.~D} {\bfseries 81} no.~10, (2010) 104019, {\ttfamily arXiv: 0912.5297 [astro-ph.CO]}.

\bibitem{Loeb}
P.~Pani and A.~Loeb, {\em JCAP} 1406 (2014) 026, {\ttfamily arXiv: 1401.3025 [astro-ph.CO]}.

\bibitem{Tinyakov}
F.~Capella {\em et~al.}, {\ttfamily arXiv: 1403.7098 [astro-ph.CO]}.

\bibitem{Tinyakov2}
G.~Defillon {\em et~al.}, {\ttfamily arXiv: 1409.0469 [gr-qc]}.

\bibitem{2000hep.ph....5271R}
S.~G.~Rubin, M.~Y.~Khlopov, and A.~S.~Sakharov, {\em Grav. Cosmol.} 6 (2000) 51, {\ttfamily arXiv: hep-ph/0005271}.

\bibitem{2011GrCo...17..181G}
A.~V.~Grobov  {\em et~al.}, {\em Grav. Cosmol.} 17 (2011) 181.

\bibitem{Landscape}
M.~R. Douglas, {\em JHEP} 5 (2003) 46.

\bibitem{random}
S.~G. Rubin,  {\em Grav. Cosmol.}, 9 (2003) 243-248.

\bibitem{random1}
S.~G. Rubin, {\ttfamily arXiv: 1403.2062 [gr-qc]}.

\bibitem{random2}
K.~A.~Bronnikov, S.~G.~Rubin, I.~V.~Svadkovsky, {\em Phys. Rev.~D} 81 (2010) 084010.







\bibitem{INTEGRAL}
G.~Weidenspointner {\em et~al.}, {\em Astron.Astrophys.} 450 (2006) 1012, {\ttfamily arXiv: astro-ph/0601673}.

\bibitem{Titarchuk}
L. Titarchuk, P. Chardonnet, {\em Astrophys. J.} 641 (2006) 293--301, {\ttfamily arXiv: astro-ph/0511333}.

\bibitem{Dolgov}
C. Bambi, A.~D. Dolgov, A.~A. Petrov, {\em Phys. Lett. B} 670 (2008) 174--178, {\ttfamily arXiv: astro-ph/0801.2786}.

\bibitem{reion}
K.~M.~Belotsky and A.~A. Kirillov, {\ttfamily arXiv: 1409.8601 [astro-ph.CO]}.

\bibitem{Ginzburg}
V.~S. Berezinsky and V.~L. Ginzburg, {\em Astrophysics of cosmic rays}, North-Holland, 1990.


\bibitem{2003A&AT...22..727D}
V.~I. Dokuchaev and Y.~N. {Eroshenko}, {\em AApTr} {\bfseries 22} (2003) 727--730, {\ttfamily arXiv: astro-ph/0209324}.

\bibitem{2008ARep...52..779D}
V.~I. Dokuchaev, Y.~N. Eroshenko, and S.~G. Rubin, {\em Astronomy Reports} {\bfseries 52} (2008) 779--789, {\ttfamily arXiv: astro-ph/0801.0885}.

\bibitem{2005GrCo...11...99D}
V.~I. Dokuchaev, Y.~N. Eroshenko, and S.~G. Rubin,  {\em Grav. Cosmol.} {\bfseries 11} (2005) 99--104, {\ttfamily arXiv: astro-ph/0412418}.

\bibitem{2011GrCo...17...27B}
K.~M. Belotsky {\em et~al.}, {\em Grav. Cosmol.}
  {\bfseries 17} (2011) 27--30.

\bibitem{2011APh....35...28B}
K.~M. {Belotsky} {\em et~al.}, {\em Astropart. Phys.} {\bfseries 35} (2011) 28--32.

\bibitem{2009ApJ...697.1071A}
W.~B. {Atwood} {\em et~al.}, {\em Astrophys. J.}
  {\bfseries 697} (2009) 1071--1102, {\ttfamily arXiv: 0902.1089 [astro-ph.IM]}.

\bibitem{2003A&A...411L...1W}
C.~Winkler {\em et~al.}, {\em Astron. \&
  Astrophys.} {\bfseries 411} (2003) L1--L6.

\bibitem{1992NASACP3137....3C}
	R.~A. Cameron {\em et~al.}, {\em NASA Conference Publication}  {\bfseries 3137} (1992) 3--14.

\bibitem{2010ApJS..188..405A}
A.~A. {Abdo} {\em et~al.}, {\em Astrophys. J. Suppl.} {\bfseries 188} (2010) 405--436, {\ttfamily arXiv: 1002.2280
  [astro-ph.HE]}.

\end{thebibliography}

\end{document}